\documentclass{emulateapj}
\usepackage{amsmath, amsthm, amssymb}

\slugcomment{Submitted to the Astrophysical Journal}

\shorttitle{OFF-AXIS ENERGY GENERATION AND THE BROAD-LINE REGION}
\shortauthors{GASKELL}

\begin{document}

\title{OFF-AXIS ENERGY GENERATION IN ACTIVE GALACTIC NUCLEI: EXPLAINING BROAD-LINE PROFILES, SPECTROPOLARIMETRIC OBSERVATIONS, AND VELOCITY-RESOLVED REVERBERATION MAPPING}

\author{C. MARTIN GASKELL}

\affil{Department of Astronomy, University of Texas, Austin, TX 78712-0259.}

\email{gaskell@astro.as.utexas.edu}

\begin{abstract}

It is shown that broad-line region (BLR) line profiles ranging from the classic ``logarithmic'' profile to double-peaked, disk-like profiles are readily explained by the distribution of BLR gas proposed by Gaskell, Klimek, \& Nazarova (GKN) without any need to invoke fundamental differences in the AGNs other than differing viewing angles.  It is argued that the highly-variable thermal energy generation in AGNs originates off axis in regions that cannot be axially symmetric.  This off-axis model readily explains the varying degrees of temporal correlation found in multi-wavelength variability studies, the strong, variable asymmetry of BLR line profiles, the varying time delays in the response of the BLR to different continuum events, how narrow velocity ranges of line profiles will often appear to respond differently or not at all to continuum variability, complex changes in the Balmer decrement with velocity, inconsistent and variable inflow/outflow signatures found in velocity-resolved reverberation mapping, the diversity of velocity-dependent polarizations observed, and polarization variability. The fundamentally non-axisymmetric nature of AGN continuum variability severely limits what can be learned from reverberation mapping.  In particular, high-fidelity reverberation mapping is not possible.  There will be systematic orientation-dependent errors in black hole mass determinations.  The effects of off-axis emission will mask subtle signatures of possible close supermassive black hole binaries.  Some tests of the off-axis-variability model are proposed.
\end{abstract}


\section{Introduction} 

The nature of the broad-line region (BLR) and the role it plays in AGNs have long been enigmatic (see reviews by \citealt{mathews+capriotti85}, \citealt{osterbrock+mathews86}, \citealt{osterbrock91}, \citealt{osterbrock93} and \citealt{sulentic+00}).  However, in recent years, a clear picture of the structure and kinematics of the BLR has emerged (see \citealt{gaskell09} for a detailed review).  The BLR is simply the inner extension of the torus \citep{netzer+laor93,suganuma+06,gaskell+07b} and it shares a similar flattened distribution and covering factor (\citealt{gaskell+07b}; hereinafter GKN).  In the GKN model the BLR is self-shielding near the equatorial plane, so there is strong ionization stratification \citep{macalpine72}, in agreement with observations (see \citealt{gaskell+07b}, \citealt{gaskell+08}, and \citealt{gaskell09}).  The covering factor of the BLR, like that of the torus, is quite large, so the appearance of the BLR is, as suggested by \citet{mannucci+92}, somewhat like a ``bird's nest''. This is illustrated in Fig.~7 of \citet{gaskell09}. Because of the dusty torus, we usually see the BLR close to face-on.  The dominant motion of the BLR is rotation, but there has to be significant vertical motion to maintain its thickness \citep{osterbrock78}. In addition to these motions there is a net inflow \citep{gaskell+goosmann08} which is providing the feeding of the black hole.

In sections 2 and 3 below I show that the GKN model readily explains classic ``logarithmic'' line profiles \citep{blumenthal+mathews75,baldwin75} and double-peaked, symmetric emission-line profiles expected from rotating disks (e.g., \citealt{shields77}), and intermediate cases, without needing to invoke fundamental differences in the AGNs.  The GKN model also reproduces observed transfer functions and velocity-delay diagrams.

Despite progress in understanding the BLR, a number of major problems have remained.  As will be discussed below, these include: (a) the strong asymmetry that can be found in the broad Balmer lines, (b) differing time delays of the BLR response to continuum events occurring close together in time, (c) a curious lack of correlation (or even an anti-correlation) between broad line variability and continuum variability over narrow velocity ranges at certain times, (d) the contradictory and un-physically variable BLR kinematics implied by velocity-resolved reverberation mapping, (e) object-to-object differences in the velocity dependence of line polarization and, (f) changes in this polarization with time.  I demonstrate in the remaining sections of the paper that all these problems and others can be readily resolved if the continuum variability is off-axis.

\section{Modelling BLR Line Profiles and Variability} 

A computer code, {\it BL-RESP}, has been developed to model BLR observations.  {\it BL-RESP} generates many observable quantities and makes movies of BLR cloud motions seen from various viewing angles.  Outputs include profiles of emission lines, and the temporal response of the BLR as a function of time delay or ``lag'' ($\tau$), flux to continuum variability (the response is normally expressed as the so-called ``transfer function'', $\Psi(\tau)$; \citealt{blandford+mckee82}). {\it BL-RESP} also gives the response as a function of projected radial velocity, $v$, to give so-called ``velocity-delay diagrams'' \citep{ulrich+horne96}.  {\it BL-RESP} does {\em not} include detailed atomic physics.  It also does not include the effects of scattering which can modify profiles (see \citealt{gaskell+goosmann08}) to produce the blueshifting of high-ionization lines \citep{gaskell82}.  The program can be used with the Monte Carlo radiative transfer code {\it STOKES} \citep{goosmann+gaskell07} which produces detailed models of line and continuum polarization due to multiple scatterings.

To match the distribution of gas in the GKN model, the BLR structure is approximated as a spherical distribution with polar bi-cones removed.  A real BLR probably consists of a fractal distribution of clouds \citep{bottorff+ferland01}, but for ease of computation {\it BL-RESP} assumes that the BLR consists of condensations (``clouds'').

Because pressure is inadequate to support the observed thickness of the BLR there must be vertical motion of the gas.  This is also required to explain line profiles and the consistency of mass estimates from the BLR.  The inflow of material onto the black hole is driven by the magneto-rotation instability (MRI) \citep{balbus+hawley91}.  Simple energy equipartition would suggest that turbulence out of the plane is comparable to the MRI turbulence in the plane of the disc and so it is highly likely that magnetic instability is also responsible for the vertical motions.  However, because there is not yet a satisfactory theory of these complexities, {\it BL-RESP} simply models the vertical structure and motions by having clouds move in tilted circular orbits.  This model assumes that BLR clouds are effectively collisionless and that their motions are dominated by gravity.  This has the advantage of involving a known force.  In reality magnetic fields are almost certainly playing a key role both in the BLR kinematics and in preventing the clouds from being destroyed in hypersonic collisions.  Magnetic fields can prevent destruction of clouds in shocks because the relevant speed in calculating the Mach number is the Alfven speed rather than the sound speed (e.g., \citealt{kennel+coroniti84}). Such considerations are, however, far beyond the scope of this paper. The distribution of tilts was taken to be uniform.

The remaining parameters are the inner and outer radii, the radial distribution of clouds (taken here to be a power-law), the viewing angle, the distance dependence of the response to the continuum, and the magnitude of the inflow component of the cloud velocity. The models do not consider how local events will modify the local ionization structure of the BLR, although {\it BL-RESP} can be used to model such refinements.

The relative sizes of the regions emitting each line and the effective range of radii over which that line is emitted are predicted by the GKN model  (see Fig.~2 of \citealt{gaskell09}). The relative sizes agree with reverberation mapping (see Fig.~1 of \citealt{gaskell+08}). Reverberation mapping can be used to set the scale factor.  In this paper detailed discussion will be restricted to the Balmer lines, for which the best data are available.  What can be inferred from other lines will be discussed elsewhere. The relative range of radii over which the Balmer lines are emitted is well-determined observationally (a) from the widths of $\Psi(\tau)$ recovered from reverberation mapping observations and (b) from line profile fitting.  These sizes are in excellent agreement with the predictions of the GKN model.   For H$\beta$, the range of radii over which the line is emitted is surprisingly narrow.  The GKN model predicts a range of a factor of four or so (see Fig.~2 of \citealt{gaskell09}), while the best $H\beta$ transfer functions also give a range of a factor of four or so in radius (see Fig.~8 of \citealt{pijpers+wanders94}).
Such a range of radius is consistent with inner and outer radii deduced by \citet{eracleous+halpern03} from fitting disk-like line profiles.  In contrast to this, for high-ionization lines like C\,IV (to be considered elsewhere), the inner radius is very small, and the range of radius is much larger (more than a factor of 10).

The typical half-opening angle of the polar cone can be inferred (a) from the covering factor implied by the observed ratio of type-1 to type-2 AGNs, and (b) from energy-balance arguments by considering the fraction of continuum radiation reprocessing by the torus and BLR (see \citealt{gaskell+07b}). The covering factor deduced from the energy-balance arguments is different for different lines, but for a line with a typical covering factor of $\sim 40$\% \citep{gaskell+07b}, it implies a half-opening angle of  60--70$^\circ$.  For C\,IV the covering factor is higher
(see \citealt{gaskell+07b}) and the half-opening angle correspondingly smaller.

The response of BLR emission at a given location to continuum variability depends on several factors including: the ionizing flux, $F_{ion}$, reaching the gas, the amount of gas at the location (which depends on the gas density and local filling factor), and details of photoionization and recombination.  The ionizing flux reaching the
gas will fall off with the distance the radiation has to travel, $r$, at least as fast as $r^{-2}$.  It will fall off faster if absorption and scattering along the line of sight are significant.  In {\it BL-RESP} the fall-off in $F_{ion}$ is parameterized as $F_{ion} \propto r^{-\alpha}$.  For the models in this paper the index, $\alpha$, was taken to be $2$.  For symmetric illumination the effect of this distance dependence of $F_{ion}$ is coupled with the effect of radial fall off in surface density of the gas.  This gives an overall emissivity power-law index, $\xi$.  This was taken to be $3$, the approximate value suggested by disk profile fitting by \citet{eracleous+halpern03}.  The results presented here are not sensitive to the precise values of $\alpha$ and $\xi$.

In this paper I use the models only (a) to show how the GKN BLR model reproduces the full-range of observed symmetric profiles, and (b) to demonstrate qualitatively how off-axis energy generation solves a wide range of major BLR puzzles.  I will not attempt to optimize parameters to match specific observations.  However, as will be shown elsewhere, it is straight forward to match observed line profiles, transfer functions, and velocity-lag diagrams, and in doing so to investigate constraints on the above parameters.

\section{BLR Line Profiles Arising from Symmetric Central Illumination} 

The Balmer lines of the majority of AGNs studied by \citet{baldwin75} showed symmetric profiles that could be well fit by a ``logarithmic'' profile \citep{blumenthal+mathews75}.   On the other hand, other AGNs show disk-like Balmer line profiles (see for example \citealt{eracleous+halpern94,strateva+03,gezari+07}).  \citet{eracleous+halpern94} suggested that AGNs with disk-like Balmer profiles are fundamentally different from AGNs with strong, symmetric, centrally-peaked profiles.  Fig.~1 shows the profiles arising from the GKN model with the basic parameters as described in the previous section.  When the BLR model is viewed from 30$^\circ$ off axis, the line profile shows the expected characteristic
double humps.   This profile agrees well with the variable part of NGC~5548 H$\beta$ profile shown in Fig.~4 of \citet{shapovalova+04}.  When {\em the same} BLR model is viewed from face-on it can be seen from Fig.~1 that a classical ``logarithmic'' profile is seen instead.  This demonstrates that disk-like line profiles can arise from the same sorts of BLRs that produce logarithmic profiles and thus supports the idea that {\em AGNs showing disk-like and logarithmic line profiles are fundamentally the same with the only difference being the viewing angle.}  Other apparent differences between AGNs with disk-like and logarithmic profiles can also be explained by differing orientation.  In particular, \citet{gaskell+04} have demonstrated that the apparent continuum differences in AGNs seen off axis are simply due to reddening.

\begin{figure}
\resizebox{\hsize}{!}{\includegraphics{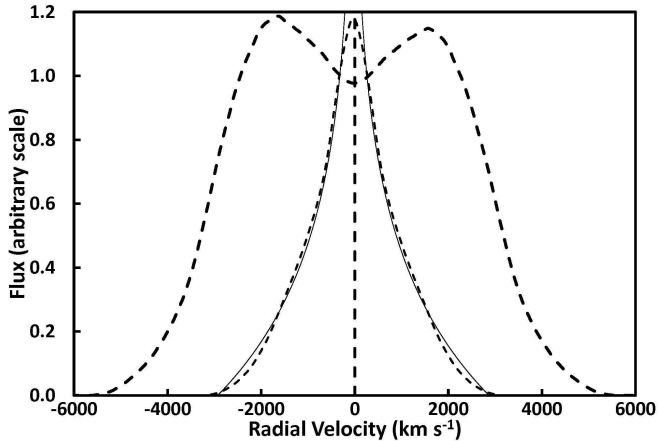}}
\caption{The dot-dashed line shows the profile that would arise from the GKN model with parameters as described in the text if it were centrally illuminated and viewed from 30$^\circ$ off axis.  The profile with short dashes is the same model viewed from face-on. The thin line superimposed on this is the logarithmic profile of \citet{blumenthal+mathews75}.}
\label{}
\end{figure}

\section{Off-Axis Continuum Variability} 

The bulk of the energy in AGNs comes out in the thermal ``big blue bump'' (see Fig.~2 of \citealt{gaskell08}).  This energy is produced by dissipative heating of matter accreting onto the black hole.  This was discussed in detail by \citet{pringle+rees72} and \citet{shakura+sunyaev73} in the context of optically-thick, but geometrically-thin, accretion disks.  With only slight modification (the addition of heat flow from the inner hotter regions to the cooler outer regions) this scenario (or quasi-spherical accretion) reproduces the time-averaged spectral energy distribution of AGNs well (see \citealt{gaskell08}).   The huge problem with steady-accretion models, however, is that, after subtraction of the host galaxy starlight contribution, the amplitude of variability of the thermal big blue bump is {\em enormous} \citep{gaskell07,gaskell08}.  Rather than AGN variability being a small perturbation of the standard disk model, the variability mechanism {\em is} the main source of energy \citep{gaskell+klimek03}.

It is pointed out in \citet{gaskell08} that this highly variable energy generation {\em must necessarily be non-axially symmetric.}  As is well known, for an accretion disk, the strongest contribution to the emission at a given frequency arises at a radius that depends on the radial temperature profile of the disk.  For a given spectral region the emission thus arises predominantly from within an annulus.  Because information cannot propagate around a ring in an accretion disk on the timescale of AGN variability (which is often as short as the light-crossing timescale), it
is clearly impossible to get an entire annulus to vary rapidly, simultaneously, and symmetrically in brightness by an order of magnitude. Instead, strong variability {\em must} arise in azimuthally- and radially-localized regions.  It will thus be highly non-axisymmetric. \citet{gaskell08} argued that the complexity found in multi-wavelength variability monitoring (see, for example, \citealt{breedt+09}) is consistent with this picture and proposed a number of tests of the off-axis energy generation scenario.

The regions of the disk with an average temperature high enough to produce ionizing photons are surprisingly close to the BLR (see Table 1 of \citealt{gaskell08}).  When there is a flare it will be occurring in a lower-temperature region which will be further out and hence closer to the Balmer-line-emitting region of the BLR or inside it.  In the following sections I consider the effects of this.

\section{Variable Lags} 

As is well known, BLR variability lags continuum variability and the lag gives an indication of the size of the BLR \citep{lyutyi+cherepashchuk72,cherepashchuk+lyutyi73}.  The simplest and most widely used way of calculating the lag is to cross-correlate line and continuum light curves \citep{gaskell+sparke86,gaskell+peterson87}.  \citet{netzer+maoz90} pointed out that different continuum events in the intensive 1989 monitoring of NGC\,5548 \citep{clavel+91,peterson+91} gave lags that differed by more than the well-understood observational errors.  This is discussed more extensively in \citet{maoz94}.  NGC~4151 shows a similar discrepancy \citep{malkov+97}.  The differences in lag (a factor of $\sim 50$\% in NGC\,5548) mean that a single transfer function will not be able to reproduce the line-flux variability corresponding to the continuum variability.  Such discrepancies are found for essentially all reverberation-mapped AGNs .

It has been pointed out \citep{gaskell08} that off-axis continuum variability readily explain differences in lags for different continuum events.  For a BLR tilted to our line of sight we see the BLR responding with less delay to a continuum flare on the far side of the black hole than to a flare on the near side of the black hole.  If the BLR is nearly face on to us, there is a similar effect if the flare is out of the plane of the BLR and towards us versus being behind the plane.

The magnitude of these effects was modeled using {\it BL-RESP}. For a line-emitting region spanning a factor of four in radius, with a half-opening angle of 65$^\circ$, tipped 30$^\circ$ relative to our line of sight (conditions approximating the H$\beta$-emitting region of the BLR of NGC\,5548), a flare at the inner edge of the BLR in the equatorial plane on the side nearest the observer produces a lag 50\% longer than a similar flare on the opposite side.  The difference increases to a factor of two for an edge-on BLR.

\newpage

\section{Variable Transfer Functions}

$\Psi(\tau)$ for H$\beta$ is usually consistent with being zero at zero time delay (e.g., \citealt{pijpers+wanders94}). As is expected for central illumination of the GKN model, this means that there is little or no gas close to the line of sight.  The observed transfer function for NGC~5548 in the 1990 observing season is shown in Fig.~2.  It can be seen that $\Psi(\tau)$ calculated for central illumination of the GKN model is a good fit.  Fig.~2 also shows $\Psi(\tau)$ for the same BLR geometry illuminated by an off-axis variable continuum source.  $\Psi(\tau)$ is shown for the source being at various azimuths.  It can be see that when the source of continuum variability is at inferior conjunction: (a) the lag is smaller (as discussed in the previous section) and (b) $\Psi(\tau)$ is predicted to be non-zero at $\tau = 0$.  When the continuum source is at superior conjunction the lag is longer and $\Psi(\tau)$ is narrower.  When the source is near either maximum elongation the difference in $\Psi(\tau)$ from the central-illumination case is smaller.

\begin{figure}
\resizebox{\hsize}{!}{\includegraphics{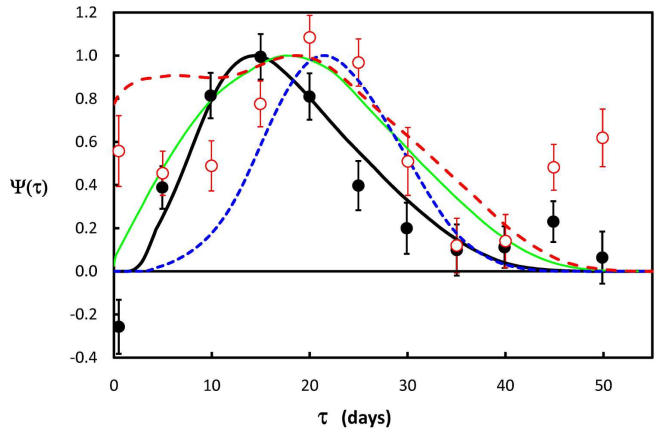}}
\caption{Observed transfer functions, $\Psi(\tau)$, as a function of lag, $\tau$, compared with $\Psi(\tau)$ calculated by {\it BL-RESP}.  The filled black circles are $\Psi(\tau)$ derived by \citet{pijpers+wanders94} for NGC~5548 for the 1990 observing season.  The solid black curve is $\Psi(\tau)$ for a centrally-illuminated model based on the GKN geometry.  The half-opening angle has been taken to be 65$^\circ$ and the H$\beta$-emitting region to cover a factor of three in radius. The system is viewed from 20$^\circ$ off-axis.  The long-dashed red curve shows $\Psi(\tau)$ for the same BLR illuminated from the inner edge of the H$\beta$-emitting region nearest the observer (at inferior conjunction).  The short-dashed blue curve shows $\Psi(\tau)$ when the source of variable illumination is at superior conjunction, and the thin green curve shows $\Psi(\tau)$ when the illumination is at either maximum elongation.  The open red circles are the observed $\Psi(\tau)$ for the 1989 observing season \citep{pijpers+wanders94}.}
\label{}
\end{figure}

As the variable source of ionization changes position one expects, one therefore expects that the shape of $\Psi(\tau)$ will change.  The only object sufficiently well observed to test this is NGC~5548.  For the years for which \citet{pijpers+wanders94} give $\Psi(\tau)$ it is similar to the 1990 one shown in Fig.~2.  The exception is the 1989 observing season where $\Psi(\tau)$ is non-zero at $\tau = 1$ unlike the other years\footnote{Note that the off-axis illumination model predicts that $\Psi(\tau)$ is poorly determined for 1989 because, as noted in Section 5, there were active regions in different locations producing different lags.  One single $\Psi(\tau)$ will not fit the entire observing season.   As explained in Section 13.1, in this situation programs to derive $\Psi(\tau)$ attempt to improve the fit to the light curves by adding in spurious structure to $\Psi(\tau)$ at large $\tau$.  $\Psi(\tau)$ is therefore uncertain at large $\tau$.  The quoted error bars for 1989 are underestimated because they do not take the change in illumination position into account}.

\section{Line Profiles} 

Disk-like, double-peaked profiles are certainly seen in many AGNs (see, for example, the studies of \citealt{eracleous+halpern94}, \citealt{eracleous+halpern03}, \citealt{strateva+03}, \citealt{gezari+07} and references therein).  \citet{gaskell+snedden99}, \citet{popovic+04}, and \citet{bon+06} have argued that such profiles are probably present in all BLRs but are simply hard to recognize when the disk is near to face-on.  However, a big problem for simple disk BLR models is that usually one peak is significantly stronger than the other \citep{gaskell83,gaskell96b,gezari+07}.  These line-profile asymmetries have been attributed to asymmetries in the BLR disks.  Suggested causes of asymmetries have included ``hot spots'' in the BLR emission \citep{zheng+91}, elliptical disks \citep{eracleous+95}, and warped disks \citep{wu+08}. Although such models can explain some line profiles, there are at least three major problems with these explanations.  The first is that Keplerian shear will quickly wipe out asymmetric distributions of BLR gas.  The second is that the line asymmetries can be very large.  Extreme examples can be seen in Fig.~1 of \citet{gaskell83} and in \citet{boroson+lauer09}.  A third problem is that profiles change with time \citep{veilleux+zheng91,zheng+91}.  These changes can be huge. For example, Mrk~668 (= OQ~208) had a strong red peak in the 1978 spectra of \citet{osterbrock+cohen79}, the 1982 spectrum of \citet{gaskell83}, and 1985-1991 spectra of \citet{marziani+93}, but it had a single-peaked H$\alpha$ profile in the 1998 spectra of \citet{gezari+07}.  Such changes are merely an extremum of BLR line-profile variability.  Line-profile changes, which can appear in periods as short as a week but can persist for years, are not obviously correlated with the continuum or emission-line flux level, and are not reverberation effects \citep{zheng+91,wanders+peterson97,peterson+99}.  Different parts of line profiles show different responses to observed continuum variability \citep{zheng+91,sergeev+00,sergeev+01,sergeev+02,shapovalova+04}. \citet{sergeev+99} pointed out that the changes they observed in the Balmer-line profiles of Mrk~6 shapes cannot be caused either by matter redistribution or light-travel-time effects.  They therefore made the important suggestion that they are probably caused by changes in the anisotropy of the ionizing continuum.

The effects on line profiles of off-axis illumination of the sort described in \citet{gaskell08} can easily be investigated with {\it BL-RESP}. Fig.~2 shows how even SDSS J0946+0139, the strongest and clearest case of an anomalous H$\beta$ profile out of the $\sim$ 9,800 low-redshift SDSS AGNs analyzed by \citet{boroson+lauer10}, can readily be explained with off-axis illumination of the GKN model with the same parameters described above.  The only difference between the model spectra in Figs.~1 and 2 is that in Fig.~2 there is additional ionizing radiation being input at the inner edge of the H$\beta$ emitting region.  This continuum ``flare'', which has approximately three times the luminosity of the rest of the disk, is taken to be at maximum elongation on the receding side of the BLR disk.

\begin{figure}
\resizebox{\hsize}{!}{\includegraphics{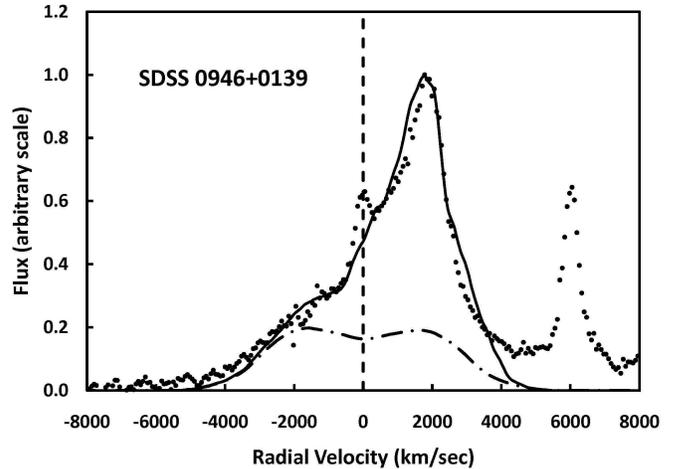}}
\caption{The observed SDSS H$\beta$ profile of SDSS J0946+0139 (dots) compared with the off-axis illumination
model described in the text (thick solid line).  The NLR contribution to H$\beta$ to the observed spectum has not been
subtracted.  The narrow emission line at +6000 km s$^{-1}$ is [O\,III] $\lambda$ 4959.  The dot-dashed line at the bottom is the same off-axis profile shown in Fig.~1 (i.e., with central illumination and viewed from 30$^\circ$ off axis).  The thick solid line shows the effect of illuminating the GKN BLR from an active region located at the inner edge of the H$\beta$-emitting region.}
\label{}
\end{figure}

\newpage

\section{Line-Profile Variability}

As noted above, line-profile variability presents difficulties for existing models of line-profile asymmetries.
In the off-axis variability model one might na\"ively expect the maximum of line-profile variability to be at the wavelength of the strongest displaced peak, but is usually observed instead to be at the edge furthest away from the line center (see, e.g., \citealt{marziani+93}).  This is due to a combination of two effects: a change in the luminosity of the off-axis, continuum-emitting region, and/or a change in position (and hence radial velocity) of the region.  Fig.~4 shows the effects of an active region orbiting close to the inner edge of the H$\alpha$ emitting region.  The basic BLR parameters are as above and the motion of the active region is as described in the figure caption.  From Fig.~4 one can see that adding the off-axis continuum emission shifts the edge of the blue peak bluewards compared to the symmetric-illumination case.  The red peak is also shifted slightly to the blue.  If the off-axis emission is combined with more symmetric illumination then the amount of shifting of the blue edge depends on the relative intensity of the active region and its orbital phase.

\begin{figure}
\resizebox{\hsize}{!}{\includegraphics{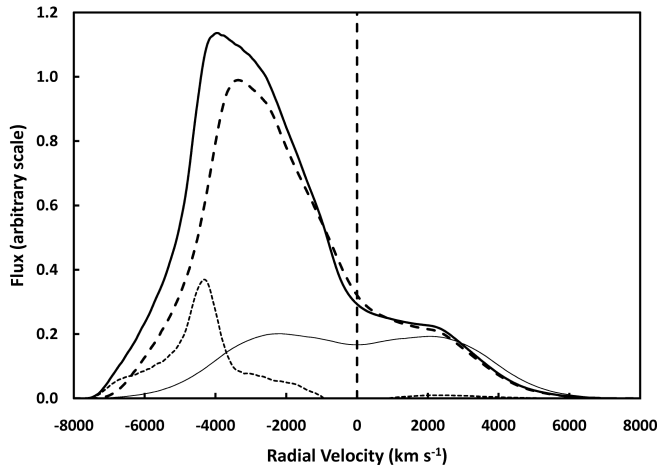}}
\caption{Profile changes due to off-axis emission at the inner edge of the H$\alpha$-emitting region of the BLR.  The thin continuous line at the bottom shows the profile expected for central illumination.  The thick solid line at the top shows the profile expected when the BLR is illuminated close to its inner edge with the active region at maximum elongation on the approaching side of the BLR.  The dashed line just below it shows how the profile changes if the active region moves 30$^\circ$ around the orbit.  The dotted line at the lower left shows the difference between the
two profiles at these two epochs.}
\label{}
\end{figure}

Since observers often report difference profiles or rms profiles, Fig.~4 shows the difference profile for the two different epochs. Although no attempt has been made to match any particular observations in detail, Fig.~4 qualitatively matches the profile variability seen in well-studied AGNs showing disk-like or highly asymmetric Balmer-line profiles.  The extensive study by \citet{gezari+07} of long-term profile variability in such objects provides many examples.  For example, Fig.~4 can be compared with the 3C~227 spectra and difference spectra in Fig.~32 of \citet{gezari+07}, and with the spectra and difference spectra for Mrk~668 shown in their Fig.~30 (if the sign of the radial velocity is reversed.)  Many further examples can be seen in the spectra of AGNs shown in the Appendix of
\citet{gezari+07}.  PKS 0235+023, for example, showed a symmetric, classic, disk-like profile in 1991, but the blue peak became much stronger by 1998 and, as predicted here, the blue peak shifted to a more negative radial velocity.

In the off-axis-variability model we expect different regions to become active at different times.  If a region remains active long enough it will change its position because of orbital motion.  The activity could also propagate inwards or outwards in the disk.  Orbital motion of regions of excess Balmer line emission was suggested by \citet{zheng+91} as the cause of line-profile variability in 3C~390.3.  \citet{gaskell96a} showed that blue peak in 3C~390.3 showed an almost linear decrease in blueshift over a period of 20 years, but this did not continue \citep{eracleous+97}.  Similar apparent orbital motion on a shorter timescale can be seen in Arp~102B (see Fig.~7 of \citealt{sergeev+00}) and later in 3C~390.3 (see Fig.~7 of \citealt{sergeev+02})

\citet{newman+07} and \citet{sergeev+00} interpreted the changing excesses of emission in the Balmer-line profiles of Arp~102B as ``hot spots'' orbiting with a period of a little over 2 years.  There are several serious problems with this.  The first is that a second ``hot spot'' appearing in Arp~102B from 2000--2005 gave a mass that differed by $\sim 30$ times the measuring error \citep{gezari+07}.  A second problem is that for other AGNs with disk-like profiles the timescales are totally inconsistent with the black hole masses estimated from the relationship between stellar velocity dispersion and black hole mass \citep{lewis+10}.  A third problem is that while excess flux in a line profile is observed at times to change between the red and blue sides of a line (as is expected if it is due to a region of excess line emission orbiting), the observed radial velocity curve can be incompatible with simple orbital motion.  Fig.~5 shows an example of this.

\begin{figure}
\resizebox{\hsize}{!}{\includegraphics{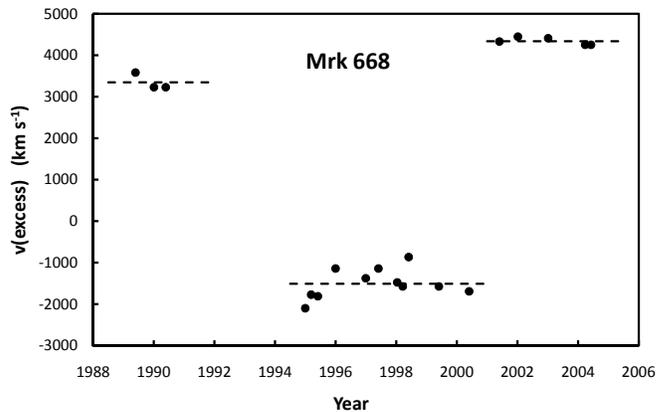}}
\caption{Radial velocity curves for regions of excess H$\alpha$ emission identified by \citet{gezari+07} in Mrk~668.  The horizontal lines are the average radial velocities for the periods indicated.}
\label{}
\end{figure}

In the off-axis variability model the excess emission is due to temporarily active regions preferentially illuminating BLR gas at a particular velocity.  An apparent change in radial velocity of excess emission, such as the changes shown in Fig.~5, is due to a region at one location becoming inactive and another region becoming active in a different location.  For high-mass black hole systems the orbital period will be long, so we expect that the radial velocity will not change slowly as an active region moves, but will change suddenly.  This sudden change in velocity seems to be what is happening in Mrk~668 (see Fig.~5).  In this case what matters is how the active regions producing the ionizing continuum radiation change, not how the structure of the BLR gas is changing.

\section{Velocity-dependent flux correlations}

In their studies of profile variability of Arp~102B, NGC~4151, and 3C~390.3 \citet{sergeev+00,sergeev+01,sergeev+02} found three curious things in all three objects that have hitherto not been satisfactorily explained:
\begin{enumerate}
\item H$\alpha$ showed strong rms variations in different velocity bins while the total line flux changed only a little.
\item The {\em integrated} line flux has a much better correlation with continuum flux than any individual velocity bin does.
\item Relatively narrow velocity regions of line profiles can at times show a strikingly weaker correlation with the continuum (or even anti-correlations.)
\end{enumerate}

All three of these effects can readily be explained by the off-axis-variability model.  Of the three the most remarkable is the lack of correlation or even anti-correlation with the continuum {\em over relatively small velocity intervals} (see especially Fig.~9 of \citealt{sergeev+01}.)  Fig.~6 illustrates where a narrow range of radial velocity has to come from.  For illustration purposes the BLR disk has been taken to be infinitely thin -- for a more typical disk scale height the region where a narrow range of velocity is coming from is more diffuse.  Different velocities in a Balmer line profile correspond to different distances of the centers of the crescents along the projected major axis of the BLR.  The narrower the velocity region, the narrower the crescent.

\begin{figure}
\resizebox{\hsize}{!}{\includegraphics{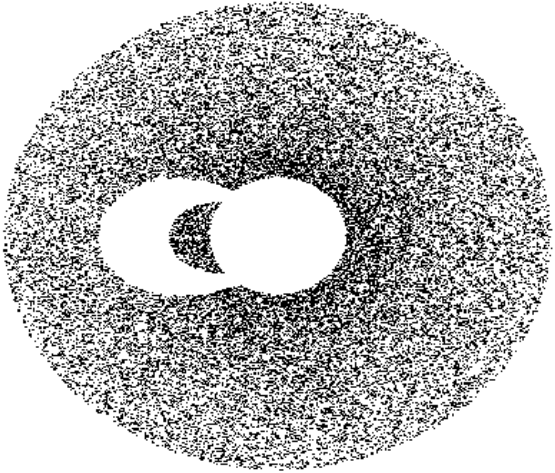}}
\caption{Schematic illustration of the Balmer-line-emitting region of an idealized, geometrically-thin BLR disk viewed from 30$^\circ$ off-axis.  The crescent-shaped region on the left shows the gas emitting in an arbitrary narrow range of observed radial velocity.  In a realistic thick BLR the boundaries of the region are not sharp.}
\label{}
\end{figure}

The first of the three effects noted by Sergeev et al. is readily explained: many different off-axis regions are contributing to the variability; each individual region will predominantly influence the line only over a narrow range of radial velocity; but the integrated flux of the line is the sum of the contributions of all the regions.  The lower rms variability of the line as a whole compared with individual velocity bins is
thus simply the result of averaging over more events.

The explanation of the second effect is similar: the observed continuum variability is the sum of all the individual off-axis events and is thus better correlated with the total line flux (which also depends on the sum of all the individual events) than with the line flux in an individual velocity bin which is dominated by the variability of a subset of the events.

The third effect arises because a minor off-axis event far out in the BLR disk might not make a detectable contribution to the total continuum flux variability seen by the observer, but it {\em will} have a strong effect on the localized region it is close to in the BLR.  At that location the variability of the closest off-axis-variability region can easily dominate over the combined variability of more distant regions.  If the local variability happens to be out of phase with the variability of the brighter, more distant, continuum-emitting regions which dominate the observed continuum light curve, the result will be to give an anti-correlation between the continuum flux and the line flux in the
velocity range the localized region is contributing to.  Conversely, if the variations are in phase the correlation will be enhanced over the velocity range.  Of these two effects, weakening the correlation will be the more obvious effect if the correlation coefficient is plotted against velocity.  Strengthening the correlation over a velocity range will be more obvious if the significance level is plotted against velocity (e.g., as in \citealt{lewis+10}).

The effect of off-axis variability on the velocity-dependent correlation coefficients has been qualitatively modeled in Fig.~7 (see figure caption for details).  Note that this is to illustrate how strong variations in the correlation coefficient can readily be produced and that no attempt has been made to input a realistic distribution of continuum events or to match the observations of any object.

\begin{figure}
\resizebox{\hsize}{!}{\includegraphics{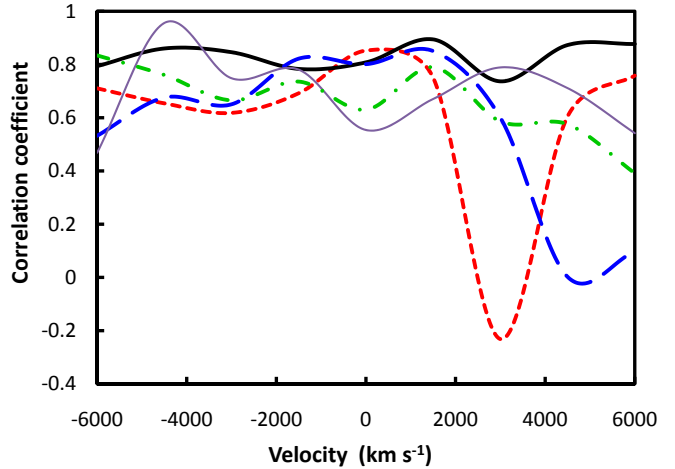}}
\caption{The effect of off-axis variability on the velocity-dependent correlation coefficients.  Continuum variability
has simply been assumed to be the result of events of different amplitudes at different radii.  Approximately half a dozen regions were taken to be active at any moment and the sampling time of the observations was taken to be longer than the lifetime of the events. The continuum flux at any instant is the sum of fluxes of all the individual events.  Each event was assumed to most strongly influence a velocity range of $\pm 1000$ km s$^{-1}$.  The distribution of events was taken to be uniform in line-of-sight velocity.  Different lines correspond to different seeds of the random number generator and show the range of curves seen.  The correlation coefficient between the continuum and the line as a whole was always greater than 0.9.}
\label{}
\end{figure}

The main thing to note from Fig~7 is that the off-axis-variability model readily produces narrow velocity regions where the line flux is not correlated with the continuum.  If we compare Fig~7 with Fig.~9 of \citet{sergeev+01} we can see that in NGC~4151 the velocity ranges of non-correlation with the continuum flux can be even narrower than produced by the off-axis-variability model in Fig.~7 (in 1997--1998, for example).  Similarly narrow regions of enhanced correlation can be found in CBS~74 (see Fig.~12 of \citealt{lewis+10}) and PKS 0921-123 (see Fig.~13 of \citealt{lewis+10}).  The narrowness of these regions implies that the region of the BLR influenced by an off-axis event can be quite small.

Three important things that analyzes such as those shown in Fig.~9 of \citet{sergeev+01} do is to establish that:
\begin{enumerate}
\item there is significant off-axis ionizing continuum variability occurring out at the radius of the Balmer-line-emitting region, and
\item the off-axis variability is not far above or below the BLR disk.
\item the Balmer line emitting region is indeed in a flat disk.
\end{enumerate}
This is because (a) the radii are accurately pinned down by the velocities at which poor line-continuum correlations are seen, and (b) neither a very thick BLR nor continuum events far from the equatorial plane will produce the sharp velocity dependence of the correlation coefficient.

Note that, in general, the spectral region correlated best with the continuum is not the region that shows the strongest variability, although in some cases it can be (e.g., at $-7500$ km s$^{-1}$ in Pictor A -- see Fig.~11 of \citealt{lewis+10}).  What matters is the phase of the local variability. For the most variable part of a line this might or might not be in phase with the dominant continuum variability.

\section{Velocity-Resolved Reverberation Mapping} 

Velocity-resolved reverberation mapping shows that the motions of BLR gas clouds are gravitationally dominated rather than arising from an outflow \citep{gaskell88,koratkar+gaskell89}.  This is because the red and blue wings vary closely in phase on a light-crossing timescale.  On average the red wing leads slightly, implying a slight net inflow (see \citealt{gaskell+goosmann08} for a summary of the case for an inflow component of motion).  While the majority of velocity-resolved reverberation mapping observations are consistent with this picture (see \citealt{gaskell10b}),
there are cases where velocity-resolved reverberation mapping clearly shows a strong signature of apparent outflow \citep{kollatschny+dietrich96,denney+09b}.  \citet{denney+09b} suggest that differing kinematic signatures reflect real object-to-object differences (e.g., a strong outflow in some AGNs). \citet{gaskell10b}, however, points out that differing kinematic signatures have been seen {\em in the same objects} at different times (e.g., \citealt{kollatschny+dietrich96,welsh+07}.)  The most spectacular example occurred in the 1989 monitoring of NGC~5548 \citep{clavel+91,peterson+91} where \citet{kollatschny+dietrich96} found strong outflow {\em and} inflow signatures for C\,IV and the Balmer lines in successive outbursts.  Since these comparable outbursts were separated by only 100 days (about the light-crossing time for the outer BLR) a real change of direction of motion of the entire BLR is out of the question.

Off-axis illumination provides a simple explanation of what is going on.  If from our viewpoint the off-axis continuum variability is displaced towards the approaching side of BLR, the blueshifted side of the line profile will vary before the redshifted side.  This gives an apparent ``outflow'' signature.  Conversely, if the continuum variability is displaced towards the receding side, the red wing variability leads the blue to give an ``inflow'' signature.  The velocity dependence of the lags is similar in both cases; only the sign of the velocity is different.  The solid
curve in Fig.~8 shows the velocity dependence of the H$\beta$ lag predicted by {\it BL-RESP} for a 4:1 ratio of inner to outer radius, a 65$^\circ$ half opening angle, a 30$^\circ$ tilt to our line of sight, and a ``flare'' at maximum elongation right at the inner edge of the BLR. The velocity dependence of the lags was also calculated for an accelerating wind and is shown as a dotted line in Fig.~7.  These theoretical curves can be compared with the four best examples of a strong velocity dependence in the H$\beta$ lag: the 1993-1995 monitoring of Mrk~6 \citep{sergeev+99}, the 2007 monitoring of NGC~3516 and NGC 3227 \citep{denney+09b}, and the 2008 monitoring of Mrk~40 \citep{bentz+08}.
NGC~3227 shows the opposite sign of the velocity dependence to the other three, so it has been plotted with the velocity scale reversed.  The four AGNs have different line widths and total lags.  To facilitate comparisons the lags of both observations and models have been normalized to a 15-day lag at zero velocity.  Choosing an appropriate normalization for the line widths is harder since different commonly used line width estimators give significantly different widths.  The velocity normalizations have therefore been adjusted to minimize residuals from the off-axis flare model (the thick solid line). Error bars in the lags have been omitted for clarity, but they can be found in the original papers.  They are comparable to the scatter about the theoretical line at low velocities, and larger at the highest velocities.

\begin{figure}
\resizebox{\hsize}{!}{\includegraphics{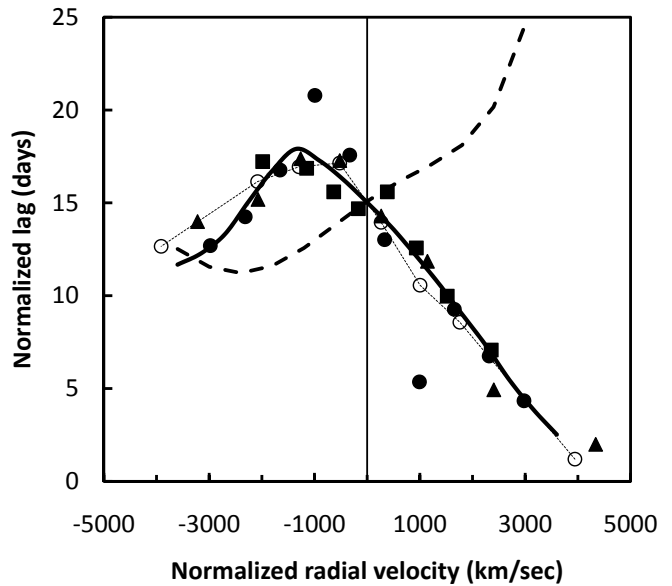}}
\caption{Observed radial velocity dependence of lags in Mrk~6 (solid circles), NGC~3516 (squares) , NGC~3227 (open circles joined by dotted line), and Mrk~40 (triangles).  The velocity scale for NGC~3227 has been reversed.  The solid line shows the predicted lags as a function of radial velocity for the GKN model with off-axis continuum variability.  The dashed line shows the predicted lags for an accelerating outflowing wind.}
\label{}
\end{figure}

The first thing to notice in Fig.~8 is that an accelerating outflowing wind model is a poor fit to the NGC~3227 lags.  In the wind model the high-velocity red wing should show the largest lag, but the observed lag decreases in the wings.  It can be seen instead that, after reversing the velocities, the velocity dependence of the NGC~3227 lags is completely consistent with the velocity dependence for the other three AGNs. This supports the idea of a common origin of the velocity dependence of the lags.  Finally, it can be seen that there is good agreement with the velocity dependence predicted by the off-axis-variability model.  Other than the scaling in velocity, there has been no attempt to optimize model parameters.  The shape of the theoretical curve is largely independent of these.  The main effect of changing parameters is to increase or decrease the amplitude of the blue/red difference.  For example, a greater tilt to the line of sight increases the difference, while moving the ``flare'' closer to the black hole, or changing to phase away from maximum elongation reduces the difference.  The off-axis variability shown in Fig.~8 produces about the maximum effect possible for the viewing angles expected for most AGNs.  This should not cause concern because the four AGNs were chosen to be extreme cases.

Although an outflow (either accelerating or constant-velocity) is excluded even for NGC~3227, infall is not excluded for the other three.  To reproduce the strong red-wing/blue-wing difference the inflow would need to be close to radial freefall (\citealt{bentz+10} have independently obtained the same result for Mrk~40 = Arp~151).  {\it BL-RESP} shows that the velocity dependence of the lag for pure infall is essentially indistinguishable from the off-axis model shown in Fig.~8.  However, pure radial infall does not explain how different kinematic signatures are seen at different times.  Mrk~6, for example, showed a much less marked red-wing/blue-wing difference in 1996-1997 \citep{sergeev+99}.
More extreme examples of changing kinematic signatures have been mentioned above.

\section{Hydrogen and Helium Line Ratios}

Although local BLR ionization perturbations caused by nearby continuum variability have not been included here there will be such effects.  In the GKN model high-ionization emission (e.g., He\,II) comes from much closer in than the Balmer line emission and hence is seen at higher velocities.  Off-axis variability, if it produces enough flux with energies above 4 Rydbergs, will temporarily cause additional enhanced He\,II emission further out in the disk (and hence at lower velocities).  The model therefore predicts that He\,II, like the Balmer lines will show anomalous variability behavior over the same narrow velocity ranges and at the same time as the Balmer lines.  This is precisely what \citet{sergeev+01} find (see their Fig.~9).

It has long been known empirically \citep{shuder82,crenshaw86} that the Balmer decrement is flatter for higher velocity BLR gas.  Photoionization modelling shows that the Balmer decrement flattens (H$\beta$/H$\alpha$ increases) with increasing ionizing photon flux \citep{kwan84,snedden+gaskell07}.  In the off-axis illumination model the photon flux is higher in the regions closest to the continuum sources.  We therefore expect that these regions will have flatter Balmer decrements.  This means that ``bumps'' in the line profiles due to off-axis illumination will be more prominent in H$\beta$ than in H$\alpha$.  Inspection of Fig.~2 of \citet{shapovalova+10} reveals a number of epochs when this is the case (e.g., in July 1996, January 1999, and January/February 2000).

\section{Line Polarization}

Polarization gives powerful insights into the structure of AGNs \citep{antonucci02}.  Polarization is extremely sensitive to departure from symmetry about the line of sight and therefore provides a powerful test of the off-axis illumination model being presented here.

Line emission from a symmetrical disk will be polarized \citep{chen+halpern90}.  Spectropolarimetric observations of Arp~102B  \citep{antonucci+96} did indeed show polarization of the line but this was inconsistent with the predictions of \citet{chen+halpern90}.  The line polarization was twice as great as predicted and, contrary to the model predictions, the polarized flux was higher in part of the blue peak than in the red peak, and it also dropped too low in the far blue wing.  The polarization of the blue side changed dramatically with wavelength, contrary to the prediction.  The polarization angle changed relative to that of the continuum by $\simeq 30^\circ$ compared with the continuum rather than the predicted $\simeq 90^\circ$.

Spectropolarimetry of AGNs in general shows a diversity of line polarization (see examples in the \citealt{smith+02} spectropolarimetric atlas).  It is normal for the polarization angle to vary across a line and also for a line to be less polarized on one side and more polarized on the other.  Differing line polarization behaviors have been interpreted as consequences of different kinematics and scattering geometries in different objects with, for example, scattering from a rotating region dominating in one object and scattering from a wind in another \citep{axon+08}.

The off-axis continuum emission discussed here produces both line and continuum polarization.  The polarization behavior is easy to understand qualitatively.  Polarization arises from a departure from symmetry about the line of sight.  A face-on disk, for example, has no asymmetry about the line of sight, and hence the integrated light has no polarization, but the polarization increases as the disk inclination is increased (see, for example, \citealt{goosmann+gaskell07}).  Off-axis illumination similarly produces strong polarization, but the polarization measured by the observer is diluted by the combined effects of illumination from other directions.  As discussed above, the gas nearest a source of off-axis illumination is brightest and shows the strongest variability.  However, because polarization measures asymmetry, the gas nearest an off-axis source of illumination will show the {\em lowest} polarization because the illumination is most symmetric nearest the source of illumination.  BLR gas on the opposite side of the disk will show the highest polarization as a result of that particular off-axis continuum source even though the illumination on the opposite side will not be particularly bright.  This means that if the source of off-axis illumination is on the receding side of the BLR, the red side will show the lowest polarization (because the illumination is relatively symmetric), while the blue side of the line will show higher polarization.  If one looks at the polarized flux, the line will appear to be blueshifted.  If the source of off-axis illumination is on the other side of the line the effects are simply reversed in velocity.

Detailed quantitative modeling needs a code such as {\it STOKES} which handles multiple-scatterings and the atomic physics of line polarization.  Such modelling will be discussed elsewhere, but since the primary effects are geometric the {\em qualitative} behavior of the polarization can be derived from {\it BL-RESP} output.  Fig.~9 shows the total line flux, the relative percentage of line polarization, and the effect on the polarization position angle of the line plus an arbitrary polarized continuum.  For illustration purposes the BLR is assumed to be excited by an active region at an azimuth 45$^\circ$ from the line of nodes in the middle of the Balmer line emitting disk.

\begin{figure}
\resizebox{\hsize}{!}{\includegraphics{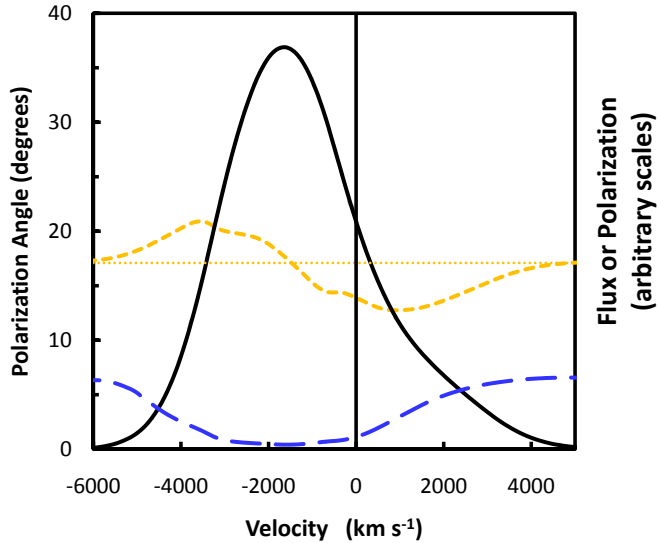}}
\caption{Total line flux (solid black line), line polarization (long-dashed blue line), and combined polarization of the line plus continuum (short-dashed burnt orange line) for a BLR illuminated by off-axis emission in the middle of the BLR at 45$^\circ$ from the line of nodes.  The light dotted line (burnt orange) is the continuum polarization position angle.  For the total flux and line polarization, zero is at the bottom of the graph but the vertical scale factor is arbitrary.}
\label{}
\end{figure}

The off-axis illumination model predicts that, in addition to this characteristic velocity dependence of the degree of polarization, there will always be a change in the angle of polarization as one goes across the line center, except in the unlikely cases of the off-axis illumination source lying exactly on the line of nodes or going through inferior or superior conjunction.  In all models where the angle of BLR polarization is at least slightly different from the angle of continuum polarization there will be a change in the position angle of the polarization at an emission line. Because the change in angle is simply the result of adding the Stokes vectors the change will be greatest at the line center. This can be seen in Fig.~9.

Spectropolarimetric observations strongly support both the predicted degree of polarization and position-angle behaviors.  The wavelength at which the excess flux is centered is indeed also the velocity at which the percentage polarization is lowest. If we go to minus that velocity on the other side of the line the polarization is high.  This can be seen in Figs.~6, 7, 10, 13, 14, 19, and 23 of \citet{smith+02} and it is especially clear in E1821+643 \citep{robinson+10}. These examples include four cases of the flux excess and polarization minimum being on the red side of the line, and three cases of them being on the blue side of the line.  The velocity dependence of the degree of  polarization shown in Fig.~9 of this paper can be directly compared with the continuum-subtracted polarization shown in Fig.~23 of \citet{smith+02}.  The S-shaped change in position angle shown in Fig.~9 as one goes across the line profile can be seen in the majority of AGNs in the \citet{smith+02} atlas (see Figs.~1ab, 2, 5b, 6, 7ab, 10, 11, 12ab, 13, 14, 15, 16, 20, 22, and 23).  More detailed analysis of individual objects will be presented elsewhere.

The alternative explanations offered in the literature for the diverse spectropolarimetric observational results are that the wavelength dependence is the result of rotation in some objects, and scattering off a wind in others.  \citep{robinson+10} have also made the suggestion that in E1821+643 the the broad-line region has two components moving with different bulk velocities away from the observer and toward a scattering region at rest in the host galaxy because of gravitational recoil due to anisotropic emission  of gravitational waves following the merger of a supermassive black hole (SMBH) binary.  In contrast to these diverse explanations in the literature, in the  model presented here, all of the spectropolarimetric observations are simply explained by one process: off-axis illumination.

\section{Discussion}

\subsection{The Limits of Reverberation Mapping}

Off-axis variability severely limits what can be achieved with reverberation mapping.  It means that goals such as trying to map possible spiral density waves in BLRs \citep{horne+04} are unattainable.  As pointed out in \citet{gaskell10b}, making further progress in reverberation mapping is not simply a matter of getting better sampling and higher signal-to-noise ratios; it is the nature of continuum variability in the AGNs and the response to it which is the fundamental limit.  No matter how good the sampling, {\em there is a strong risk of getting very misleading results from observing campaigns of short duration because the results are dominated by a small number of events.}  It is more important to have {\em longer} monitoring campaigns rather than denser sampling. In the past a lag has been considered to be well-determined if one complete event (a clear rise and fall or vice versa) is well sampled in both the continuum and line light curves.  However, we have seen here that because of off-axis variability, lags from single events are untrustworthy.  Even two or three successive events could give a misleading result because, as we have seen, active regions can persist for much longer than the light-crossing time.  Unfortunately the need to observe more than a single complete event adds to what is already a very labor intensive process.

The off-axis-variability model means that we have to look in a new way at variability data in general and reverberation mapping in particular.  Events have to be looked at {\em individually}.  Attempts to reconstruct $\Psi(\tau)$ have often looked at as long a series of observations as possible.  Since the lines respond differently to different events, a single transfer function will fit some events poorly.  A fitting routine that minimizes residuals by adjusting $\Psi(\tau)$ will attempt to improve a poor fit to the line response by adding in a response to a much earlier event (i.e., at larger $\tau$.)  For methods that force $\Psi(\tau)$ to be positive (methods such as the maximum entropy method used by \citealt{krolik+91}) this will give spurious features at large $\tau$ and an erroneous impression that the BLR is larger than it really is.  Examples of such features can be seen in the transfer functions of C\,IV and Mg\,II for NGC~5548 shown in Fig.~11 of \citet{krolik+91}. On close inspection it can be seen that even with the addition of such features the line light curves systematically fail to follow the convolution of the observed continuum curve with $\Psi(\tau)$ (e.g., around MJD 7540).

To learn from reverberation mapping it is important to look at line profiles as well as integrated fluxes.  For planning future reverberation mapping campaigns, good spectral resolution and high signal-to-noise ratios in lines are needed.  Although the sort of ``high-fidelity'' mapping envisioned by \citet{horne+04} is unattainable, reverberation mapping, if properly interpreted, still tells us a lot about the structure of the BLR and about the nature of continuum variability. A good illustration of what can be accomplished is the recent study of \citet{jovanovic+10} who track the movement of the dominant region of off-axis variability in 3C~390.3 observed in 1995--1999.  In agreement with what is found here, they find that the off-axis emission dominates over axisymmetric emission.


\subsection{Black Hole Masses}

Off-axis variability has multiple implications for the estimation of black hole masses from the BLR \citep{dibai77}. To get masses from the BLR we need to know that the BLR is gravitationally bound.   One then needs to know an effective radius, an effective velocity, and the appropriate scaling factor.  The gravitational domination of the BLR was established with velocity-resolved reverberation mapping \citep{gaskell88}, but, as discussed above, some velocity-resolved reverberation mapping \citep{kollatschny+dietrich96,denney+09b} has given contradictory results.  The good news is that apparent outflow signatures from velocity-resolved reverberation mapping are the result of off-axis variability rather than outflows. The bad news is that because of the dominance of off-axis variability, reverberation times (our most direct way of getting effective radii) cannot be trusted when only a single event is observed (see above).

\citet{dibai77} introduced a simpler method of estimating the effective radius of the BLR from the AGN luminosity.  Because of its ease of use, the Dibai method has now been applied to tens of thousands of AGNs.  \citet{bochkarev+gaskell09} show that even the earliest mass estimates by the Dibai method agree well with reverberation mapping and argue that this is because of the similarity of continuum shapes, BLR geometries, and BLR kinematics.  This results in the tight relationship between BLR radius and continuum luminosity. Much of the scatter in this relationship must be consequence of off-axis variability.  As the quality of reverberation mapping data improve, the tightness of the relationship is therefore expected to improve (see Fig.~6 of \citealt{denney+10}).  However, in the absence of adequate reverberation mapping coverage of a given AGN it is preferable to estimate the black hole mass by the Dibai method.

It has been common in recent years to estimate the effective velocity of BLR gas from rms spectra.  The justification has been that the width of the rms spectrum is a better indicator of the velocity of the variable component of the BLR gas than the width of average spectra is.  The off-axis-variability model says instead that the rms spectrum will have lumps and spikes in it because of the multiple off-axis variability events and hence that it will be a {\em poorer} indicator of the effective velocity of the BLR than the average spectrum.  Lumps and spikes are clearly observed in many rms spectra (see for example, the strong blue spike in the 2007 rms spectrum of NGC~5548 in  \citealt{denney+10}). The main use of rms spectra is in removing non-varying narrow-line region contributions.


Assuming that the GKN model is an approximately correct description of the BLR, the potentially most serious problem in determining masses is the orientation effect.  It can be seen from Fig.~1 that for the range of inclinations over which the BLR can be readily observed in the optical (0 to $30^\circ$) the observed FWHM (the most commonly used indicator of the effective BLR velocity) varies by about a factor of three for the same BLR.  Since virial mass estimates depend on the square of the velocity, the black hole mass is being overestimated by an order of magnitude for highly-inclined systems compared with face-on ones.  The inclination effect thus needs to be allowed for in estimating masses. The magnitude of the effect depends on the thinness of the BLR.  The best indicator of the inclination of the BLR is the flatness of the peak of the line profile after removing the narrow-line contribution.

\subsection{Looking for Signatures of Sub-parsec Supermassive Binary Black Holes}

It is now agreed that massive galaxies (ellipticals and disk galaxies with classical bulges) grow through mergers.  The central black holes of merging galaxies will form binary black holes \citep{begelman+80}. \citet{gaskell83} suggested that close supermassive binary black holes could be detected through their effect on the BLR, and, in particular, that displaced broad line peaks could be a consequence of the orbital motions of close binary black holes.  Reverberation mapping of 3C~390.3, however, showed that on a light-crossing timescale the red and blue peaks varied near simultaneously \citep{dietrich+98,obrien+98}.  This strongly rules out binary black holes as the cause of the displaced peaks \citep{gaskell+snedden99,gaskell10a}.  Nonetheless, binary black holes certainly must form, and because of their importance in the evolution of black holes and galaxies, increasing attention has been devoted to finding examples of close (sub-parsec scale) binaries.  \citet{boroson+lauer09} have recently suggested that the asymmetric displaced BLR peaks in SDSS J1536+0441 are due to a binary black hole system (i.e., as in \citealt{gaskell83}.)  \citet{chornock+10} and \citet{gaskell10a} argue that SDSS J1536+0441 is simply an extreme example of an AGN with asymmetric, disk-like emission instead.  The demonstration here (see Fig.~2) of how easily off-axis emission can produce extreme profiles (J1536+0441 is less extreme than the example in Fig.~2) adds strong support to J1536+0441 having a normal BLR.

The recent debate over J1536+0441 does raise the interesting and important question of whether sub-parsec supermassive black hole binaries {\em can} be detected through anomalous BLR profiles.  Because off-axis emission can readily produce complex profiles, the answer is unfortunately almost certainly ``no''.

\subsection{Other Possible Causes of Non-axisymmetric Emission}

Although accretion disk variability must necessarily be off-axis and although, as discussed above, such variability successfully explains a wide range of hitherto puzzling BLR observations, there are a number of other plausible causes of non-axisymmetric variability.
\begin{enumerate}

\item \citet{gaskell+klimek03} point out that the short timescale of the strong variability of AGNs requires the transmission of large amounts of energy in particles with relativistic or near-relativistic energies. Such motions will result in anisotropic emission (beaming).  \citet{gaskell06} proposed that the combination of correlated and uncorrelated short-timescale, multi-wavelength variability in AGNs could be a consequence of such anisotropic high-energy emission (see Fig.~5 of \citealt{gaskell06}.)

\item Anisotropy and variability can also be due to blocking of radiation from the inner regions of the AGN \citep{gaskell10b}.  There is abundant evidence now for variable X-ray absorption, often with partial coverage, in the inner regions of AGNs (see \citealt{turner+miller09}, \citealt{gaskell10b} and references therein).  \citet{gaskell10b} points out that this would attenuate radiation hitting the BLR in certain directions.

\item Another thing that could block radiation from the center is warping of a thin accretion disk \citep{kinney+00}.  Such warping is well-known in Galactic X-ray binaries such as SS~433 and Her~X-1 (see \citealt{ogilvie+dubus01} for discussion) and there are a number of possible mechanisms which could produce a similar warping in AGNs \citep{kinney+00}.  \citet{bentz+10} have recently shown that an alternative explanation of the apparent strong inflow signature in Mrk~40 could be anisotropic illumination of the BLR due to warping of the inner accretion disk.

\end{enumerate}

\subsection{Unanswered Questions}

Although the combination of the geometry of the GKN BLR model and off-axis variability has strong support and is successful in explaining many observations of line and continuum variability, there are many unanswered questions.  These include:
\begin{enumerate}
 \item What is the mechanism producing the strong variability?
 \item What is the radial and vertical distribution of the regions of strong variability (``flares'')?
 \item What is the energy distribution of flares? (and how does it depend on radius?)
 \item What governs how long a continuum-emitting region is active?
 \item What maintains the vertical motions of the BLR gas?
 \item What happens when rapidly moving regions of higher density BLR gas (``clouds'') collide?  Why are the clouds not destroyed?
 \item How much damage is done to the BLR disk by the strong variability close to it?
\end{enumerate}

Although a lot of further observational and theoretical study is needed, there are indications of some possible answers.  For example, we can infer that the flares are not far above the BLR because it would be hard to produce the observed frequent lack of a line/continuum variability correlation over a narrow velocity range (see Section 9).  If the flares are high above the BLR their effect would be spread out over too large a range of velocity.  We can also say that although AGN variability is not azimuthally symmetric on short timescales, if we average over a long enough time it is expected to be symmetric.  Just how long is ``long enough'' is not clear.  For some AGNs, asymmetric emission has persisted for as long as the AGN has been observed.

For the radial distribution of activity, there certainly is some randomness, but if energy generation is averaged over a sufficiently long time period, the radial dependence of the energy production must match that expected for a steady accretion flow (see \citealt{gaskell08}).  The frequency of structure in line profiles as a function of velocity and time can reveal information about the radial distribution of continuum variability.  \citet{flohic+eracleous08} have already deduced that there is an apparent increase in small-scale structure in the outer part of the H$\beta$ emitting region.  They interpret this as increased clumping in the outer BLR due to gravitational instabilities.  In the off-axis variability scenario this structure is due to small scale variability in the continuum rather than clumping of the BLR.

\subsection{Predictions}

The off-axis illumination model presented here makes many testable predictions.  The strongest of these, and the easiest to test, are the polarization predictions since polarization is highly sensitive to departures from symmetry about the line of sight.

As discussed in Section 12, the off-axis illumination model predicts that there will be a minimum in the polarization near the location of the continuum source.  The velocity of excess emission in a line profile will be the velocity of the minimum in the line polarization.  This can be checked by looking at the polarization of AGNs showing asymmetric broad lines.  In contrast to the predictions of the off-axis illumination model, if a warp in the disk causes only part of the outer disk to be centrally illuminated (as suggested by \citealt{bentz+10}), the illuminated side of the disk will be highly polarized and the polarization will peak at the peak of the line emission. This would give a quite different wavelength dependence of the polarization from that predicted by the off-axis illumination model.

The most important tests come from polarization {\em variability}.  As different regions in different locations become active the general level of polarization and its velocity dependence will change.  Changes in the velocity dependence of polarization have already been observed (e.g., in NGC 4151; \citealt{martel98}).  The off-axis-illumination model predicts that the velocity dependence of the percentage of polarization will reverse when the excess flux in a line changes sides.  If there is a small change in the position of the main region of illumination due to orbital motion or other drift of the active region, this will show up as a change in position angle of the polarization and in the velocity dependence of this angle.  The best candidates for testing these prediction are AGNs where the orbital timescale is not too long and there has been a history of the relative intensities of red and blue excesses switching.

Light echoes in broad-band polarized flux have already been reported (see \citealt{gaskell+07a}).  Measuring velocity-dependent spectropolarimetric lags should be a powerful tool for distinguishing between models of the structure and kinematics of the BLR and continuum.   Although it is beyond the scope of the present paper, it is straight forward to use {\it BL-RESP} and {\it STOKES} to predict time-dependent and velocity-dependent spectropolarimetric variations for the off-axis variability model and other models.

The off-axis variability model makes predictions about the velocity dependence and variability of line ratios (see Section 10).  Bumps due to off-axis illumination should be stronger in H$\beta$ than in H$\alpha$.  He\,I and He\,II should show the same anomalous responses over the same narrow velocity ranges as the hydrogen lines.

In checking the model presented here there is obviously a need to make everything consistent (e.g., changing lags, changing velocity dependence, line profiles, spectropolarimetry).  Note though that things will not always be quite as simple as might na\"ively be expected.  For example, the brightest region of off-axis-illumination might not be the most variable during a given time period.

\section{Conclusions}

Although we do not understand the physical mechanism causing AGN variability, it must be highly non-axisymmetric. This is a major paradigm shift in our understanding of AGNs.  It has been argued in \citet{gaskell08} that off-axis continuum emission explains many puzzles of AGN continuum variability, and it is argued here that such variable emission also readily explains a wide variety of hitherto puzzling observations of AGNs.  Things explained include:
\begin{enumerate}
\item the different lags measured for different events in reverberation-mapping campaigns,
\item the unusually asymmetric line profiles seen in some AGNs,
\item line-profile variability (including sharp structures seen in difference and rms spectra),
\item the strange line flux and continuum correlations or non-correlations often seen over remarkably small velocity intervals,
\item the sometimes rapidly changing and conflicting kinematic signatures seen in velocity-resolved reverberation mapping,
\item why variable bumps in line profiles are more pronounced in H$\beta$ than H$\alpha$,
\item velocity-dependent line polarization
\item line-polarization variability

\end{enumerate}

An important related conclusion is that a single BLR geometry with the structure advocated by \citet{gaskell+07b}, and with kinematics consistent with this, successfully explains a wide variety of BLR observations once the presence of off-axis variability is recognized.  AGNs with very different line profiles are not fundamentally different.  In particular, {\em AGNs showing disk-like line profiles are not fundamentally different from AGNs showing classical logarithmic profiles.}

It is clear that orientation has a significant effect on line profiles.  This will introduce systematic errors in  black hole mass estimates.

While more exotic explanations of the various phenomena considered here are not necessarily ruled out (e.g., binary black holes, recoiling black holes, or warped accretion disks), the off-axis variability model renders them unnecessary.  Off-axis variability also makes it very hard to detect such phenomena from observations of broad lines.

The complications created by off-axis variability need to be taken into account when interpreting reverberation-mapping results and planning monitoring campaigns.  These complications severely limit what can be learned from reverberation mapping.  In particular the goal of detecting fine structure in BLRs seems unattainable.

\acknowledgments

I would like to express my appreciation to the organizers and participants of several recent AGN conferences which provided valuable stimuli for this work: the 2007 meeting in Huatulco, Mexico, the 2008 meeting on Crete, the 2009 meeting in Zrenjanin, Serbia, and the 2009 meeting in Como, Italy.  I am also grateful to Ski Antonucci, H\'el\`ene Flohic, Ren\'e Goosmann, Pawan Kumar, Milo\v{s} Milosavljevi\'c, and Ed Robinson, for useful discussions.  This research has been supported through NSF grants AST 03-07912 and AST 08-03883, and NASA grant NNH-08CC03C.


\begin{thebibliography}{}

\bibitem[\protect\citeauthoryear{Antonucci}{2002}]{antonucci02} Antonucci, R.~R.~J.\ 2002, in Astrophysical Spectropolarimetry, eds. J. Trujillo-Bueno, F. Moreno-Insertis, \& F. S\'anchez (Cambridge:                    Cambridge Univ. Press), p.\ 151

\bibitem[\protect\citeauthoryear{Antonucci et al.}{1996}]{antonucci+96} Antonucci, R., Hurt, T., \& Agol, E.\ 1996, \apjl, 456, L25

\bibitem[\protect\citeauthoryear{Axon et al.}{2008}]{axon+08} Axon, D.~J., Robinson, A.,
Young, S., Smith, J.~E., \& Hough, J.~H.\ 2008, Mem. Soc. Astron. Italiana, 79, 1213

\bibitem[\protect\citeauthoryear{Baldwin}{1975}]{baldwin75} Baldwin, J.~A.\ 1975, \apj, 201, 26

\bibitem[\protect\citeauthoryear{Balbus \& Hawley}{1991}]{balbus+hawley91} Balbus, S.~A. \& Hawley, J.~F 1991, ApJ, 376, 214

\bibitem[\protect\citeauthoryear{Begelman, Blandford, \& Rees}{2000}]{begelman+80} Begelman, M.~C., Blandford, R.~D. \& Rees, M.~J.\ 1980, Nature, 287, 307

\bibitem[\protect\citeauthoryear{Bentz et al.}{2008}]{bentz+08} Bentz, M.~C. et al.\ 2008, ApJ Lett, 689, L21

\bibitem[\protect\citeauthoryear{Bentz et al.}{2010}]{bentz+10} Bentz, M.~C. et al. 2010, ApJ in press [arXiv:1007.0781]

\bibitem[\protect\citeauthoryear{Blandford \& McKee}{1982}]{blandford+mckee82} Blandford, R.~D., \& McKee, C.~F.\ 1982, \apj, 255, 419

\bibitem[\protect\citeauthoryear{Blumenthal \& Mathews}{1975}]{blumenthal+mathews75} Blumenthal, G.~R., \& Mathews,
W.~G.\ 1975, ApJ, 198, 517

\bibitem[\protect\citeauthoryear{Bochkarev \& Gaskell}{2009}]{bochkarev+gaskell09} Bochkarev, N.~G., \& Gaskell, C.~M.\ 2009, Ast.\@ Lett., 35, 287

\bibitem[\protect\citeauthoryear{Bon et al.}{2006}]{bon+06} Bon, E., Popovi{\'c}, L.~{\v C}., Ili{\'c}, D., \& Mediavilla, E.\ 2006, New Astron. Rev., 50, 716

\bibitem[\protect\citeauthoryear{Boroson \& Lauer}{2009}]{boroson+lauer09} Boroson, T.~A., \& Lauer, T.~R. 2009, Nature, 458, 53

\bibitem[\protect\citeauthoryear{Boroson \& Lauer}{2010}]{boroson+lauer10} Boroson, T.~A., \& Lauer, T.~R. 2010, AJ, 140, 390

\bibitem[\protect\citeauthoryear{Bottorff \& Ferland}{2001}]{bottorff+ferland01} Bottorff, M. \& Ferland, G.~J. 2001, ApJ, 549, 118

\bibitem[\protect\citeauthoryear{Breedt et al.}{2009}]{breedt+09} Breedt, E., et al.\ 2009, \mnras, 394, 427

\bibitem[\protect\citeauthoryear{Chen \& Halpern}{1990}]{chen+halpern90} Chen, K., \& Halpern, J.~P.\ 1990, \apjl, 354, L1

\bibitem[\protect\citeauthoryear{Chen et al.}{1989}]{chen+89} Chen, K., Halpern, J.~P.,
\& Filippenko, A.~V.\ 1989, \apj, 339, 742

\bibitem[\protect\citeauthoryear{Cherepashchuk \& Lyutyi}{1973}]{cherepashchuk+lyutyi73}
Cherepashchuk, A. M. \& Lyutyi V. M. 1973, Ap. Lett, 13, 165

\bibitem[\protect\citeauthoryear{Chornock et al.}{2010}]{chornock+10} Chornock, R., et al.\ 2010, \apjl, 709, L39

\bibitem[\protect\citeauthoryear{Clavel et al.}{1991}]{clavel+91} Clavel, J., et al.\ 1991, \apj, 366, 64

\bibitem[\protect\citeauthoryear{Crenshaw}{1986}]{crenshaw86} Crenshaw, D.~M.\ 1986, ApJS, 62, 821

\bibitem[\protect\citeauthoryear{Denney et al.}{2009a}]{denney+09a} Denney, K.~D., et al.\ 2009a, ApJ, 702, 1353

\bibitem[\protect\citeauthoryear{Denney et al.}{2009b}]{denney+09b} Denney, K.~D., et al.\ 2009b, ApJ Letts. 704, L80

\bibitem[\protect\citeauthoryear{Denney et al.}{2010}]{denney+10} Denney, K.~D., et al.\ 2010, ApJ. in press [arXiv:1006.4160]

\bibitem[\protect\citeauthoryear{Dibai}{1977}]{dibai77} Dibai, \'{E}. A. 1977, Soviet Astron.\@ Lett., 3, 1

\bibitem[\protect\citeauthoryear{Dietrich et al.}{1998}]{dietrich+98} Dietrich, M., et al.\ 1998, ApJS, 115, 185

\bibitem[\protect\citeauthoryear{Eracleous \& Halpern}{1994}]{eracleous+halpern94} Eracleous, M., \& Halpern, J.~P.\ 1994, ApJS, 90, 1

\bibitem[\protect\citeauthoryear{Eracleous et al.}{1997}]{eracleous+97} Eracleous, M., Halpern, J.~P., Gilbert, A.~M., Newman, J.~A., \&
Filippenko, A.~V.\ 1997, \apj, 490, 216

\bibitem[\protect\citeauthoryear{Eracleous \& Halpern}{2003}]{eracleous+halpern03}  Eracleous, M., \& Halpern, J.~P.\ 2003, ApJ, 599, 886

\bibitem[\protect\citeauthoryear{Eracleous et al.}{1995}]{eracleous+95} Eracleous, M., Livio, M., Halpern, J. P., \& Storchi-Bergmann, T. 1995,
ApJ, 438, 610

\bibitem[\protect\citeauthoryear{Flohic \& Eracleous}{2008}]{flohic+eracleous08} Flohic, H.~M.~L.~G. \& Eracleous, M.\ 2008, \apj, 686, 138

\bibitem[\protect\citeauthoryear{Gaskell}{1982}]{gaskell82} Gaskell, C.~M.\ 1982, ApJ, 263, 79

\bibitem[\protect\citeauthoryear{Gaskell}{1983}]{gaskell83} Gaskell, C.~M.\ 1983, Liege International Astrophys. Colloq., 24, 473

\bibitem[\protect\citeauthoryear{Gaskell}{1988}]{gaskell88} Gaskell, C.~M.\@ 1988, ApJ, 325, 114

\bibitem[\protect\citeauthoryear{Gaskell}{1996a}]{gaskell96a} Gaskell, C.~M.\ 1996a, ApJ. Lett., 464, L107

\bibitem[\protect\citeauthoryear{Gaskell}{1996b}]{gaskell96b} Gaskell, C.~M. 1996b, in Jets from Stars and
Galactic Nuclei, ed. W. Kundt. (Berlin: Springer-Verlag), p. 165

\bibitem[\protect\citeauthoryear{Gaskell}{2006}]{gaskell06} Gaskell, C.~M.\ 2006, Astron. Soc. Pacific Conf. Ser., 360, 111

\bibitem[\protect\citeauthoryear{Gaskell}{2007}]{gaskell07} Gaskell, C.~M.\ 2007, in The Central Engine of Active Galactic Nuclei, ed. L. C. Ho \& J.-M. Wang (San Francisco: ASP), ASP Conf. Ser. 373, 596

\bibitem[\protect\citeauthoryear{Gaskell}{2008}]{gaskell08} Gaskell, C.~M.\ 2008, Rev. Mexicana Astron. Ap. Conf. Ser., 32, 1

\bibitem[\protect\citeauthoryear{Gaskell}{2009}]{gaskell09} Gaskell, C.~M. 2009, New Astron. Rev, 53, 140

\bibitem[\protect\citeauthoryear{Gaskell}{2010a}]{gaskell10a} Gaskell, C.~M. 2010a, Nature, 463, E1

\bibitem[\protect\citeauthoryear{Gaskell}{2010b}]{gaskell10b} Gaskell, C.~M. 2010b, in Accretion and Ejection in AGNs: a Global View, eds. L. Maraschi, G. Ghisellini, R. Della Ceca \& F. Tavecchio, ASP Conf. Ser, 427 [arXiv:0910.3945]

\bibitem[\protect\citeauthoryear{Gaskell et al.}{2004}]{gaskell+04} Gaskell, C.~M., Goosmann, R.~W., Antonucci, R.~R.~J., \& Whysong, D.~H.\ 2004, ApJ, 616, 147

\bibitem[\protect\citeauthoryear{Gaskell \& Goosmann}{2008}]{gaskell+goosmann08} Gaskell, C.~M. \& Goosmann, R.~W.\ 2008, ApJ,~submitted [arXiv:0805.4258]

\bibitem[\protect\citeauthoryear{Gaskell et al.}{2007a}]{gaskell+07a} Gaskell, C.~M., Goosmann, R.~W., Merkulova, N.~I., Shakhovskoy, N.~M., Shoji, M.\ 2007a, ApJ submitted [arXiv:0711.1019]

\bibitem[\protect\citeauthoryear{Gaskell et al.}{2008}]{gaskell+08} Gaskell, C.~M., Goosmann, R.~W., \& Klimek, E.~S.\ 2008, Mem. Soc. Ast. Italiana, 79, 1090

\bibitem[\protect\citeauthoryear{Gaskell \& Klimek}{2003}]{gaskell+klimek03} Gaskell, C.~M. \& Klimek, E.~S. 2003, Astron. Ap. Trans., 22, 611

\bibitem[\protect\citeauthoryear{Gaskell et al.}{2007b}]{gaskell+07b} Gaskell, C.~M., Klimek, E.~S., \&
Nazarova, L.~S.\ 2007b, ApJ, submitted (GKN) [arXiv:0711.1025]

\bibitem[\protect\citeauthoryear{Gaskell \& Peterson}{1987}]{gaskell+peterson87} Gaskell, C.~M.~\& Peterson, B.~M.~1987, ApJS, 65, 1

\bibitem[\protect\citeauthoryear{Gaskell \& Snedden}{1999}]{gaskell+snedden99} Gaskell, C.~M., \& Snedden, S.~A.\ 1999, in Structure and Kinematics of Quasar Broad Line Regions, eds. C.~M. Gaskell,  W.~N. Brandt, M. Dietrich, D. Dultzin-Hacyan, \& M. Eracleous (San Francisco: Astron. Soc. Pacific), ASP Conf. Ser, 175, 157

\bibitem[\protect\citeauthoryear{Gaskell \& Sparke}{1986}]{gaskell+sparke86} Gaskell, C.~M.~\& Sparke, L.~S.~1986, ApJ, 305, 175

\bibitem[\protect\citeauthoryear{Gezari, Halpern, \& Eracleous}{2007}]{gezari+07} Gezari, S., Halpern, J. P., \& Eracleous, M. 2007, ApJS, 169, 167

\bibitem[\protect\citeauthoryear{Goosmann \& Gaskell}{2007}]{goosmann+gaskell07} Goosmann, R.~W., \& Gaskell, C.~M.\
2007, A\&Ap, 465, 129

\bibitem[\protect\citeauthoryear{Horne et al.}{2004}]{horne+04} Horne, J., Peterson, B.~M., Collier, S.~J.,
\& Netzer, H. 2004, PASP, 116, 465

\bibitem[\protect\citeauthoryear{Jovanovi{\'c} et al.}{2010}]{jovanovic+10} Jovanovi{\'c}, P., Popovi{\'c}, L.~{\v C}., Stalevski, M., \& Shapovalova, A.~I.\ 2010, \apj, 718, 168

\bibitem[\protect\citeauthoryear{Kennel \& Coroniti}{1984}]{kennel+coroniti84} Kennel, C.~F., \& Coroniti, F.~V.\ 1984, \apj, 283, 694

\bibitem[\protect\citeauthoryear{Kinney et al.}{2001}]{kinney+00} Kinney, A.~L., Schmitt, H.~R., Clarke, C.~J., Pringle, J.~E., Ulvestad, J.~S. \& Antonucci, R.~R.~.J.\ 2001, ApJ, 537, 152

\bibitem[\protect\citeauthoryear{Kollatschny}{2003}]{kollatschny03} Kollatschny, W.\ 2003, A\&Ap, 407, 461

\bibitem[\protect\citeauthoryear{Kollatschny \& Dietrich}{1996}]{kollatschny+dietrich96} Kollatschny, W. \& Dietrich, M.\ 1996, A\&A, 314, 43

\bibitem[\protect\citeauthoryear{Koratkar \& Gaskell}{1989}]{koratkar+gaskell89}
Koratkar, A.~P. \& Gaskell, C.~M. 1989, ApJ, 345, 637

\bibitem[\protect\citeauthoryear{Krolik et al.}{1991}]{krolik+91} Krolik, J.~H. et al.\@ 1991, ApJ, 371, 541

\bibitem[\protect\citeauthoryear{Kwan}{1984}]{kwan84} Kwan, J.\ 1984, ApJ, 283, 70


\bibitem[\protect\citeauthoryear{Lewis et al.}{2010}]{lewis+10} Lewis, K.~T., Eracleous, M., \& Storchi-Bergmann, T. 2010, ApJS, 187, 416.

\bibitem[\protect\citeauthoryear{Lyutyi \& Cherepashchuk}{1972}]{lyutyi+cherepashchuk72}
Lyutyi, V.\,M.\ \& Cherepashchuk, A.\,M. 1972, Astron. Tsirk, 688, 1

\bibitem[\protect\citeauthoryear{MacAlpine}{1972}]{macalpine72} MacAlpine, G.~M.\ 1972, ApJ, 175, 11

\bibitem[\protect\citeauthoryear{Malkov et al.}{1997}]{malkov+97} Malkov, Y.~F., Pronik, V.~I., \& Sergeev, S.~G.\ 1997, \aap, 324, 904

\bibitem[\protect\citeauthoryear{Mannucci et al.}{1992}]{mannucci+92} Mannucci, F., Salvati, M., \& Stanga, R.~M.\ 1992, ApJ, 394, 98

\bibitem[\protect\citeauthoryear{Maoz}{1994}]{maoz94} Maoz, D.\ 1994, in Reverberation Mapping of the Broad-Line Region in Active Galactic Nuclei, eds. P.~M.\ Gondhalekar, K.~Horne, and B.~M.\ Peterson (San Francisco: Astron.\ Soc.\ Pacific), ASP Conf. Ser. 69, 95

\bibitem[\protect\citeauthoryear{Martel}{1998}]{martel98} Martel, A.~R.\ 1998, ApJ, 508, 657

\bibitem[\protect\citeauthoryear{Marziani et al.}{1993}]{marziani+93} Marziani, P., Sulentic, J. W., Calvani, M., Perez, E., Moles, M., \& Penston, M.
V. 1993, ApJ, 410, 56

\bibitem[\protect\citeauthoryear{Mathews \& Capriotti}{1985}]{mathews+capriotti85} Mathews, W.~G., \& Capriotti, E.~R.\ 1985, Astrophysics of Active Galaxies and Quasi-Stellar Objects, (Mill Valley: University Science Books) p.\ 185

\bibitem[\protect\citeauthoryear{Netzer \& Laor}{1993}]{netzer+laor93} Netzer, H. \& Laor, A.\ 1993, ApJ Lett, 404, L51

\bibitem[\protect\citeauthoryear{Netzer \& Maoz}{1990}]{netzer+maoz90} Netzer, H. \& Maoz, D. 1990, ApJ Lett, 365, L5.

\bibitem[\protect\citeauthoryear{Newman et al.}{1997}]{newman+07} Newman J.A., Eracleous M., Filippenko A.V., \& Halpern, J.P., 1997, ApJ
485, 570

\bibitem[\protect\citeauthoryear{O'Brien et al.}{1998}]{obrien+98} O'Brien, P.~T., et al.\ 1998, ApJ, 509, 163

\bibitem[\protect\citeauthoryear{Ogilvie \& Dubus}{2001}]{ogilvie+dubus01} Ogilvie, G.~I. \& Dubus, G. 2001, MNRAS, 320, 485

\bibitem[\protect\citeauthoryear{Osterbrock}{1978}]{osterbrock78} Osterbrock, D.~E.\ 1978,
Proc. Nat. Acad. Sci., 75, 540

\bibitem[\protect\citeauthoryear{Osterbrock}{1991}]{osterbrock91} Osterbrock, D.~E.\ 1991, Rep. Prog. Phys., 54, 579

\bibitem[\protect\citeauthoryear{Osterbrock}{1993}]{osterbrock93} Osterbrock, D.~E.\ 1993, ApJ, 404, 551

\bibitem[\protect\citeauthoryear{Osterbrock \& Cohen}{1979}]{osterbrock+cohen79}  Osterbrock, D.~E. \& Cohen, R.~D. 1979, MNRAS, 187, 61P

\bibitem[\protect\citeauthoryear{Osterbrock \& Mathews}{1986}]{osterbrock+mathews86} Osterbrock, D.~E., \& Mathews, W.~G.\ 1986, ARA\&A, 24, 171

\bibitem[\protect\citeauthoryear{Peterson et al.}{1991}]{peterson+91} Peterson B.~M., et al.\@1991, ApJ, 368, 119

\bibitem[\protect\citeauthoryear{Peterson et al.}{1999}]{peterson+99} Peterson B.~M., Pogge R.~W., \& Wanders I.\ 1999, in Structure and Kinematics of Quasar Broad-Line Regions, eds. C.\,M. Gaskell, W.\,N. Brandt, M. Dietrich, D. Dultzin-Hacyan, \& M. Eracleous (San Franciso: Astron. Soc. Pacific), ASP Conf.\ Ser.\ 175, 41
%

\bibitem[\protect\citeauthoryear{Pijpers \& Wanders}{1994}]{pijpers+wanders94} Pijpers, F.~P., \& Wanders, I.\ 1994, MNRAS, 271, 183

\bibitem[\protect\citeauthoryear{Popovi{\'c} et al.}{2004}]{popovic+04} Popovi{\'c}, L.~{\v C}., Mediavilla, E., Bon, E., \& Ili{\'c}, D.\ 2004, A\&Ap, 423, 909

\bibitem[Pringle \& Rees(1972)]{pringle+rees72} Pringle, J.~E., \& Rees, M.~J.\ 1972, \aap, 21, 1

\bibitem[\protect\citeauthoryear{Robinson et al.}{2010}]{robinson+10} Robinson, A., Young,
S., Axon, D.~J., Kharb, P., \& Smith, J.~E.\ 2010, \apjl, 717, L122

\bibitem[\protect\citeauthoryear{Sergeev et al.}{1999}]{sergeev+99} Sergeev, S.~G., Pronik, V.~I., Sergeeva, E.~A., \& Malkov, Yu.~F. 1999, ApJS, 121, 159

\bibitem[\protect\citeauthoryear{Sergeev et al.}{2000}]{sergeev+00} Sergeev, S.~G., Pronik, V.~I., \& Sergeeva, E.~A.\ 2000, A\&A 356, 41

\bibitem[\protect\citeauthoryear{Sergeev et al.}{2001}]{sergeev+01} Sergeev, S.~G., Pronik, V.~I., \& Sergeeva, E.~A.\ 2001, ApJ, 554, 245

\bibitem[\protect\citeauthoryear{Sergeev et al.}{2002}]{sergeev+02} Sergeev, S.~G., Pronik, V.~I., Peterson, B.~M., Sergeeva, E.~A., \& Zheng, W.\ 2002, ApJ, 76, 660

\bibitem[\protect\citeauthoryear{Shakura \& Sunyaev}{1973}]{shakura+sunyaev73} Shakura, N.~I., \& Syunyaev, R.~A.\ 1973, \aap, 24, 337

\bibitem[\protect\citeauthoryear{Shapovalova et al.}{2001}]{shapovalova+01} Shapovalova, A.~I. et al.\ 2001, A\&A, 376, 775

\bibitem[\protect\citeauthoryear{Shapovalova et al.}{2004}]{shapovalova+04} Shapovalova, A.~I. et al.\ 2004, A\&A, 422, 925

\bibitem[\protect\citeauthoryear{Shapovalova et al.}{2010}]{shapovalova+10} Shapovalova, A.~I. et al.\ 2010, A\&A, 509, 106

\bibitem[\protect\citeauthoryear{Shields}{1977}]{shields77} Shields, G.~A.\ 1977, Ap.\ Lett., 18, 119

\bibitem[\protect\citeauthoryear{Shuder}{1982}]{shuder82} Shuder, J.~M.\ 1982, ApJ, 259, 48

\bibitem[\protect\citeauthoryear{Smith et al.}{2002}]{smith+02} Smith, J.~E., Young, S., Robinson, A., Corbett, E.~A., Giannuzzo, M.~E., Axon, D.~J., \& Hough, J.~H.\ 2002, \mnras, 335, 773

\bibitem[\protect\citeauthoryear{Snedden \& Gaskell}{2007}]{snedden+gaskell07} Snedden, S.~A. \& Gaskell, C.~M.\ 2007, ApJ, 669, 126

\bibitem[\protect\citeauthoryear{Strateva et al.}{2003}]{strateva+03} Strateva, I.~V., et al. \ 2003, \aj, 126, 1720

\bibitem[\protect\citeauthoryear{Suganuma et al.}{2006}]{suganuma+06} Suganuma, M., et al.\  2006, ApJ, 639, 46

\bibitem[\protect\citeauthoryear{Sulentic et al.}{2000}]{sulentic+00} Sulentic, J.~W., Marziani, P., \& Dultzin-Hacyan, D.\ 2000, ARA\&A, 38, 521

\bibitem[\protect\citeauthoryear{Turner \& Miller}{2009}]{turner+miller09} Turner, T.~J. \& Miller, L.\ 2009, Astron. Ap. Rev., 17, 47

\bibitem[\protect\citeauthoryear{Ulrich \& Horne}{1996}]{ulrich+horne96} Ulrich, M.-H., \& Horne, K.\ 1996, MNRAS, 283, 748

\bibitem[\protect\citeauthoryear{Veilleux \& Zheng}{1991}]{veilleux+zheng91} Veilleux, S., \& Zheng, W. 1991, ApJ, 377, 89

\bibitem[\protect\citeauthoryear{Wanders \& Peterson}{1997}]{wanders+peterson97} Wanders, I. \& Peterson, B. M. 1996, ApJ, 466, 174

\bibitem[\protect\citeauthoryear{Welsh et al.}{2007}]{welsh+07} Welsh, W.~F., Martino, D.~L., Kawaguchi, G., \& Kollatschny, W.\ 2007, in The Central Engine of Active Galactic Nuclei, ed. L. C. Ho \& J.-M. Wang (San Francisco: ASP), ASP Conf. Ser., 373, 29

\bibitem[\protect\citeauthoryear{Wu et al.}{2008}]{wu+08} Wu, S.-M., Wang, T.-G., \& Dong, X.-B.\ 2008, \mnras, 389, 213

\bibitem[\protect\citeauthoryear{Zheng et al.}{1991}]{zheng+91} Zheng, W., Veilleux, S., \& Grandi, S. A. 1991, ApJ, 381, 41

\end{thebibliography}
\end{document}